\documentclass[12pt]{article}
\usepackage{amsmath}
\usepackage{graphicx}
\usepackage{url} 
\usepackage{soul}
\usepackage{amsmath,amssymb,bm}
\usepackage{changepage}
\usepackage[utf8x]{inputenc}
\usepackage{textcomp,marvosym}
\usepackage{nameref,hyperref}
\usepackage[right]{lineno}
\usepackage{microtype}
\DisableLigatures[f]{encoding = *, family = * }
\usepackage[table]{xcolor}
\usepackage{array}

\usepackage{subfigure}
\usepackage{caption}

\usepackage{amsmath,comment}
\usepackage{amssymb,amsmath}
\usepackage{graphicx}
\usepackage{enumerate}
\usepackage{natbib}
\usepackage{url} 
\usepackage{physics}

\setcitestyle{round}
\usepackage{amsthm}
\usepackage{amsfonts}
\usepackage{array}
\renewcommand{\bar}{\overline}
\renewcommand{\hat}{\widehat}
\usepackage{todonotes}
\usepackage{graphicx}
\usepackage{relsize}
\usepackage{bm}
\renewcommand{\hat}{\widehat}

\usepackage{caption}
\usepackage{multirow}
\usepackage{tabularx}
\usepackage{algorithm}
\usepackage[noend]{algpseudocode}
\usepackage{float}
\usepackage{booktabs}
\usepackage{xcolor}
\usepackage{cancel}
\usepackage{color}

\graphicspath{plot/}
\usepackage{amsmath}
\DeclareMathOperator*{\argmin}{arg\,min}
\usepackage{natbib}

\newtheorem{theorem}{\textsc{Theorem}}

\newcommand{\blind}{0}

\begin{document}

\def\spacingset#1{\renewcommand{\baselinestretch}%
{#1}\small\normalsize} \spacingset{1}


\if0\blind
{
  \title{\bf Multidimensional heterogeneity learning for count value tensor data with applications to field goal attempt analysis of NBA players}
  \author{Guanyu Hu\thanks{
    Department of Statistics, University of  Missouri - Columbia}, 
    Yishu Xue\thanks{
    Google Inc}, 
    and 
    Weining Shen\thanks{
    Department of Statistics, University of California, Irvine}}
  \maketitle
} \fi

\if1\blind
{
  \bigskip
  \bigskip
  \bigskip
  \begin{center}
    {\LARGE\bf Title}
  \maketitle
    
\end{center}
  \medskip
} \fi

\bigskip
\begin{abstract}
We propose a multidimensional tensor clustering approach for studying how professional basketball players' shooting patterns vary over court locations and game time. Unlike most existing methods that only study continuous-valued tensors or have to assume the same cluster structure along different tensor directions, we propose a Bayesian nonparametric model that deals with count-valued tensors and projects the heterogeneity among players onto tensor dimensions while allowing cluster structures to be different over directions. Our method is fully probabilistic; hence allows simultaneous inference on both the
number of clusters and the cluster configurations. We present an efficient posterior sampling method and establish the large-sample convergence properties for the posterior distribution. Simulation studies have demonstrated an excellent empirical performance of the proposed method. Finally, an application
to shot chart data collected from 191 NBA players during the 2017–2018 regular season is
conducted and reveals several interesting insights for basketball analytics.
\end{abstract}

\noindent%
 
\spacingset{1.45}

\section{Introduction}
There is a rapid growth in sports analytics over the recent years thanks to the fast development in game tracking technologies \citep{albert2017handbook}. New statistics and machine learning methods are largely needed to address a range of questions from game outcome prediction to player performance evaluation \citep{cervone2014pointwise,franks2015characterizing, cervone2016multiresolution}, and to in-game strategy planning \citep{fernandez2018wide,sandholtz2019measuring}. 

Our focus in this paper is to study the National Basketball Association (NBA) players' {\it shooting patterns}, in particular, how they change over shooting locations and game time (e.g., first quarter versus clutch time). In professional basketball research, shooting pattern (or field goal attempt) is a fundamental metric since it often provides valuable insights to players, coaches and managers, e.g., players and coaches will be able to obtain a better understanding of their current shooting choices and hence develop further offensive/defensive plans accordingly, while managers can make better data-informed decisions on player recruitment. In the literature, it is common to employ spatial and spatial-temporal models  \citep{reich2006spatial,miller2014factorized, jiao2019bayesian} to study the spatial and temporal dependence in the field goal attempt data. The key novelty of our paper is to introduce a new Bayesian tensor multidimensional clustering method that studies the heterogeneity of shooting patterns among players. 

Our starting point is to recognize that the field goal attempt data in basketball games enjoys a natural {\it tensor structure}. For example, we can divide the basketball half court into regions under the {\it polar coordinate} system and then summarize the number of field goal attempts over each region during each of the four quarters of the game as entries in a three-way tensor, where each tensor direction corresponds to the shooting distance, angle, and game time (quarter). One advantage of considering a tensor representation is that the spatial-temporal dependence structure is automatically considered as part of the tensor structure. Compared to the most existing works that rely on Cartesian coordinate system
\citep{miller2014factorized,jiao2019bayesian,yin2020bayesian,hu2020bayesiangroup,yin2020analysis1}, our approach certainly makes more sense since shooting angle and distance are two important factors that affect the shooting selection of
professional players \citep{reich2006spatial}. Studying the change of shooting patterns over time is also meaningful, e.g., Stephen Curry has made a comparable fewer number of attempts in fourth quarter during regular season simply because Gold State
Worriers has often established a significant lead at the end of the third quarter during the
2017-2018 regular season.

Tensor models have received a great deal of attention in machine learning and statistics literature \citep{kolda2009tensor,sun2014tensors,bi2021tensors}. For tensor clustering problems, most existing work \citep{sun2019dynamic,chi2020provable,mai2021doubly}  either works with a single tensor or assumes that the clustering structure is the same across different tensor directions. It is also common that they only consider tensors with entries that takes continuous or binary values. In terms of model estimation, clustering is often based on solving a regularized  optimization problem, which requires pre-specifying the number of clusters or choosing the cluster number based on certain {\it ad hoc} criteria. There are some recent papers on Bayesian tensor models, e.g., \citet{guhaniyogi2017bayesian,spencer2019bayesian,guhaniyogi2014bayesian}. However, all of them are in the regression context, hence cannot be directly applied to solve our problem. Our proposed approach differs with aforementioned methods in several ways. First, we consider a flexible multidimensional tensor clustering problem, which allows different clustering structures over tensor directions. We believe this is a meaningful relaxation in many applications, e.g., basketball players' shooting choice may differ significantly in terms of shooting distance, angle and game time  depending on players' position, shooting preference and role in the team. Moreover, we focus on count-valued tensors (rather than continuous-valued tensors) for the obvious reason that the number of shot attempts is the main outcome of interest in our application. Thirdly, our model is fully probabilistic, which allows an easier interpretation compared to optimization-based method. In particular, we consider a Bayesian nonparametric model under the mixture of finite mixtures framework
\citep[MFM;][]{miller2018mixture}, which allows simultaneous estimation and inference for the number of clusters and the associated clustering
configuration for each direction (e.g. distance, angle, and quarter). We show that the posterior inference can be conducted efficiently for our method and demonstrate its excellent empirical performance via simulation studies. We also establish posterior convergence properties for our method. Finally, our proposed method reveals several interesting data-driven insights of
field goal attempts data which are useful for professional basketball players, coaches, and
managers.

\textbf{Main contributions and broad impact:} (1) We are among the first (to the best of our knowledge) to introduce {\it tensor models} to sports analytics. It is our hope that this paper can contribute to promoting more use of tensor methods in different sport applications. (2) We develop a novel multidimensional tensor clustering approach that allows different clustering structures over tensor directions and handles count-valued data. The proposed method is fully Bayesian, which renders convenient inference on the number of clusters and the clustering structure. (3) We provide a large-sample theoretical justification for our method by showing posterior consistency for the cluster number and contraction rate for the mixing distributions. These results are new for Bayesian tensor models.


\section{Method}\label{sec:method}
\subsection{Probabilistic multi-dimensional tensor clustering }
We treat the shooting chart data as three-way tensors and discuss a
multi-dimensional clustering approach in this section. Note that each direction of the tensor represents the shooting angle, distance to the basket, and one of the four quarters in the game.  Our proposed method can
be conveniently extended to study general multi-way tensor data as well.
Let~$\bm{Y}$ be a~$p_1 \times p_2 \times p_3$ tensor with each element~$Y_{ijk}$
only taking count values for $i=1,\ldots,p_1; j=1,\ldots,p_2; k=1,\ldots,p_3$.
It is natural to consider a Poisson distribution with a mean parameter
represented as a rank-one tensor, that is, 
\begin{align}\label{eq1}
\bm{Y} \sim \text{Poisson}( 
\bm{\gamma}_{1} \circ
\bm{\gamma}_{2} \circ \bm{\gamma}_{3}), 
\end{align}
where $\circ$ denotes the outer product between two vectors,
and $\bm{\gamma}_{1} \in  \mathbb{R}_{+}^{p_1},\bm{\gamma}_{2} \in 
\mathbb{R}_{+}^{p_2},\bm{\gamma}_{3} \in  \mathbb{R}_{+}^{p_3}$. 
Model \eqref{eq1} can also be viewed as a Poisson regression model where the
mean parameter corresponds to an analysis of variance (ANOVA) model with
main effects only, that is, $\log \text{E}(Y_{ijk}) = \log \gamma_{1,i} +
\log \gamma_{2,j} +  \log \gamma_{3,k}$ for $1\leq i \leq p_1, 1\leq j
\leq p_2, 1 \leq k \leq p_3$. By ignoring the interaction effects,
the number of parameters is effectively reduced from $p_1 p_2 p_3$ to
$(p_1 + p_2+ p_3)$. Hence it renders parsimonious parameter estimation
and easy interpretation in our NBA application study, i.e., the main effects
$\log \bm{\gamma}_1, \log \bm{\gamma}_2, \log \bm{\gamma}_3$ correspond
to the additive effect of shooting distance, angle, and game time (quarter). 

In order to learn the multidimensional heterogeneity pattern, we consider
three independent  
mixture of finite mixtures \citep[MFM;][]{miller2018mixture} priors on
$\bm{\gamma}_1, \bm{\gamma}_2, \bm{\gamma}_3$ such that the clustering
pattern in those three directions can be learned separately. Here we
present a brief introduction to MFM without getting into more details.
Given $n$ observations, we consider $z_1,\ldots,z_n$ as their clustering
labels, e.g., $z_1 = z_2 = z_4$ would mean that observations $1,2,4$ belong
to the same cluster. Then the MFM prior can be expressed as 
\begin{eqnarray}\label{eq:MFM}
	K \sim p(\cdot), \quad (\pi_1, \ldots, \pi_K) \mid K  \sim \mbox{Dir} (\gamma,
	\ldots, \gamma), \quad z_i \mid K, \pi \sim \sum_{h=1}^K  \pi_h \delta_h,\quad
	i=1, \ldots, n, 
\end{eqnarray}
where $K$ is the number of clusters, $(\pi_1,\ldots,\pi_K)$ are associated
cluster weights and $\sum_{h=1}^K  \pi_h \delta_h$ is the mixture distribution
with $\delta_h$ being a point-mass at~$h$. Under the Bayesian framework, all
those three quantities are random and hence are  assigned with prior
distributions, i.e., we use 
$p(\cdot)$, which is a proper probability mass function on~$\mathbb{N}_+$,
as a prior on $K$, and a Dirichlet distribution on the mixture weights.
Compared to the Chinese restaurant process (CRP), the probability of
introducing a
new table (cluster) for MFM is slowed down by a factor of ~$V_n(t+1)/ V_n(t)$,
which allows for a
model-based pruning of the tiny extraneous clusters.
Here the coefficient~$V_n(t)$ is defined as
\begin{align*} 
	\begin{split}
	V_n(t) &= \sum_{n=1}^{+\infty}\dfrac{k_{(t)}}{(\gamma k)^{(n)}} p(k),
	\end{split}         
\end{align*} 
where $k_{(t)}=k(k-1)\ldots(k-t+1)$, and $(\gamma k)^{(n)} = {\gamma k} (\gamma
k+1)\ldots(\gamma k+n-1)$, and $\gamma$ is the hyperparameter in the Dirichlet prior for the weights. The conditional distributions of $z_i, i=2, \ldots,
n$ under~\eqref{eq:MFM} can be defined in a P\'{o}lya urn scheme similar to
CRP:
\begin{eqnarray}\label{eq:mcrp}
	P(z_{i} = c \mid z_{1}, \ldots, z_{i-1})  \propto   
	\begin{cases}
	\abs{c} + \gamma  , &  \text{at an existing table labeled}\, c\\
	V_n(t+1)/ V_n(t)\gamma,  &  \text{if}~ \, c\,
	\text{~is a new table}
	\end{cases},
\end{eqnarray}
with~$t$ being the number of existing clusters. 

Now back to the shooting chart data, as we propose to use three
independent MFM priors
for clustering shooting distance, angle, and game time, our final model
can be presented in the following hierarchical structure, 
\begin{equation}\label{eq:hierarchical_model}
	\begin{split}
	& K_{\ell} \stackrel{\text{i.i.d.}}{\sim} p_{K},  ~~ \ell=1,2,3,\\
& \bm{\pi}_{\ell} = (\pi_{\ell, 1},\ldots,\pi_{\ell,K_{\ell}}) \mid K_{\ell}
\sim
\text{Dir}(\nu,\ldots,\nu),~~, \nu > 0,~ \ell=1,2,3, \\  
&
\log{\bm{\gamma}_{\ell,1}},\ldots,\log{\bm{\gamma}_{\ell,K_{\ell}}}\stackrel
{\text{i.i.d.}}{\sim
}\text{MVN}_{p_{\ell}}(\bm{0},\Sigma_{\ell}),~~ \ell=1,2,3,\\
&\Sigma_{\ell}=\sigma^2_{\ell}(I_{\ell}-\rho_{\ell}\bm{W}_{\ell}),~~
\ell=1,2,3,\\
&\sigma^2_{\ell}\sim \text{Gamma}(a,b),~~ \ell=1,2,3,\\
&\rho_{\ell}\sim\text{Unif}(c_{1\ell},c_{2\ell}),~~ \ell=1,2,3,\\
&P(z_{i\ell} =j\mid \bm{\pi}_{\ell},K_\ell)=\pi_{j\ell},~~ \ell=1,2,3,
~~j=1,\ldots,K_\ell,\\
&\bm{Y}_i \sim \text{Poisson}( 
\bm{\gamma}_{1,z_{i1}} \circ
\bm{\gamma}_{2,z_{i2}} \circ \bm{\gamma}_{3,z_{i3}}),~i=1,\ldots,n,
	\end{split}
\end{equation} 
where the main effects for distance, angle, and period are modeled by
multivariate normal distributions and their covariance matrices involve
adjacency matrices, denoted by $\bm{W}_l$'s, that are used for incorporating
the potential spatial and temporal correlation information (e.g., two shooting 
locations are next to each other). To ensure those covariance matrices
$\Sigma_{\ell}$ are positive definite, we introduce $c_{1\ell}$ and
$c_{2\ell}$ as the reciprocals of minimum and
maximum eigenvalues of $\bm{W}_l$, respectively. For the prior $p_{K}$ on
the number of clusters, we consider a truncated
Poisson$(1)$ following the recommendations in  \citet{miller2018mixture}
and \citet{geng2019probabilistic}.

Our multidimensional clustering model in \eqref{eq:hierarchical_model} sits between two extremes. One is the usual tensor clustering model that assumes the same cluster structure across different directions, which certainly is more restrictive compared to ours. The other is to marginally cluster over each of the tensor directions and solve multiple clustering problems independently, which does not fully utilize the tensor structural information. Our method combines the attractive features from both sides by allowing cluster structures to be different over directions while  borrowing information to improve the estimation efficiency. 

Our model in \eqref{eq:hierarchical_model} 
can be viewed as a Bayesian mixture of rank-one tensor models. Compared
to the frequenstist work on tensor clustering
\citep{sun2019dynamic,chi2020provable},
where a Tucker decomposition is usually utilized and the choice of the
rank relies heavily on pre-specification or certain model selection criteria,
our approach is capable of automatically determining the rank while quantifying the uncertainty in rank selection. Moreover, our method is fully probabilistic; hence each mixture component is easy to interpret in practice.


\subsection{Theoretical Properties}\label{theory}
Next we study the theoretical properties for the posterior distribution obtained from model \eqref{eq:hierarchical_model}. For convenience, we define three mixing measures $G_{\ell} = \sum_{i=1}^{K_{\ell}} \pi_{\ell,i} \delta(\gamma_{\ell,i})$  for $\ell=1,2,3$, where $\delta(\cdot)$ is the point mass measure. In other words, $G_1,G_2,G_3$ represent the clustering structures and associated parameters along each of the three directions in the tensor. In order to establish the posterior contraction results, we consider a refined parameter space $\bm{\Theta^*}$ defined as
$\cup_{k_1,k_2,k_3=1}^{\infty} \bm{\Theta_k^*}$ for $\bm{k} = (k_1,k_2,k_3)$, where $\bm{\Theta_k^*}$ is a compact parameter space for all model parameters including mixture weights and main effects given a fixed cluster number for each direction, i.e., $K_1=k_1, K_2 = k_2,$ and $K_3 = k_3$. More precisely, we define $\bm{\Theta_k^*}$ as
\begin{align*}
     \Big\{&\pi_{\ell,i} \in (\epsilon, 1-\epsilon) ~~\text{for every}~i=1,\ldots,k_{\ell}, \ell=1,2,3, ~\sum_{j}^{k_{\ell}} \pi_{\ell,j} = 1 ~~\text{for every}~\ell=1,2,3, \\
    &   \gamma_{\ell,i} \in (-M,M)  ~~\text{for every}~i=1,\ldots,k_{\ell}, \ell=1,2,3,  \Big\},
\end{align*}
where $\epsilon$ and $M$ are some pre-specified positive constants. For any two mixing measures $G_1 = \sum_{i=1}^k p_i \delta(\gamma_i)$ and $G_2 = \sum_{j=1}^{k'} p_j' \delta(\gamma_j)$, we define their Wasserstein distance as $W(G_1,G_2) = \inf_{q \in \mathcal{Q}} \sum_{i,j} q_{ij}  |\gamma_i - \gamma_j | $, where $\mathcal{Q}$ denotes the collection of joint discrete distribution on the space of $\{1,\ldots,k\} \times \{1,\ldots,k'\}$ and $q_{ij} $ is the probability being associated with $(i,j)$-element and it satisfies the constraint that $\sum_{i=1}^k q_{ij} = p_j'$ and $\sum_{j=1}^{k'} q_{ij} = p_i$, for every $i=1,\ldots,k$ and $j=1,\ldots,k'$. 

For $\ell = 1,2,3$, let $K_{\ell}^0$ and $G_{\ell}^0$ be the true number of clusters and true mixing measure along direction $\ell$. Also let $P_0$ be the associated joint probability measure. Then the following theorem establishes the posterior consistency and contraction rate for the cluster number and mixing measure. The proof is given in the Supplementary Material; and it is based on the results for Bayesian mixture models in \citet{guha2021posterior}.  
\begin{theorem}\label{thm1}
Let $\Pi_n(\cdot \mid \bm{Y}_1,\ldots,\bm{Y}_n)$ be the posterior distribution obtained from \eqref{eq:hierarchical_model} given i.i.d. observations $\bm{Y}_1,\ldots,\bm{Y}_n$. Suppose that the true parameters belong to $\bm{\Theta^*}$. Then for each of $\ell =1,2,3$, we have \begin{align*}
\Pi_n\left\{K_{\ell} = K_{\ell}^0 \mid \bm{Y}_1,\ldots,\bm{Y}_n)\right\} \rightarrow 1, ~\text{and}~~ \Pi_n \left\{(W(G_{\ell},G_{\ell}^0)\lesssim (\log n/n)^{-1/2} \mid \bm{Y}_1,\ldots,\bm{Y}_n)\right\} \rightarrow 1, 
\end{align*}
almost surely under $P_0$ as $n \rightarrow \infty$. 
\end{theorem}
Theorem \ref{thm1} shows that as sample size $n \rightarrow \infty$, our proposed Bayesian model is capable of correctly identifying the unknown number of clusters along each of the tensor directions with posterior probability tending to one. Moreover, the latent clustering structure (e.g., cluster membership) can also be consistently recovered. The contraction rate for $G_{\ell}$ is nearly parametric with an additional logarithmic factor, which is common in the Bayesian asymptotic literature \citep{guha2021posterior}. The assumption of a compact parameter space $\bm{\Theta^*}$ is needed to rule out extreme scenarios, for example, when some mixture probabilities are extremely close to $0$, for which it becomes very challenging to distinguish between our model and a sub-model without these small mixture components. In practice, this assumption is often satisfied since we can always restrict the modeling parameters to take values within a pre-specified range, e.g., assuming cluster probability to be at least $\epsilon$ for some small $\epsilon$ values such as $.0001\%$. Our results can also be extended to general multi-way tensors as long as the independent MFM priors are used for each direction. 


\subsection{Bayesian Inference}\label{sec:inference}
We discuss the posterior sampling scheme for our model. For the MFM prior, we use the stick-breaking \citep{sethuraman1994constructive}
 approximation to reconstruct $$K_{\ell} \sim p_{K}, \bm{\pi}_{\ell} =
 (\pi_{1\ell},\ldots,\pi_{K_{\ell} \ell}) \mid K_{\ell} \sim
 \text{Dir}(\nu,\ldots,\nu),~~ \ell=1, 2, 3,$$ as follows for each of $\ell=1,2,3$,  
\begin{itemize}
\item \textbf{Step 1.} Generate $\eta_1,\eta_2,\cdots \overset{\text{iid}}{\sim}
\text{Exp}(\psi_\ell)$,
	\item \textbf{Step 2.} Let $K_\ell=\min\{j:\sum_{k=1}^j \eta_k \geq 1\}$,
	\item \textbf{Step 3.} Set $\pi_{h\ell}=\eta_h$, for $h=1,\cdots,K_\ell-1$,
	\item \textbf{Step 4.} Set $\pi_{h\ell}=1-\sum_{h=1}^{K_\ell-1}\pi_h$,
\end{itemize}
where we choose $(K_\ell-1) \sim \mbox{Poisson}(\psi_\ell)$ and $\nu=1$. Based on the stick-breaking reparameterization, we obtain a similar hierarchical model as the 
Dirichlet process mixture model in \citet{ishwaran2001gibbs} when we choose a
sufficiently large dimension $T$ for $\bm{\pi}_{\ell}$ and set the last
$T-K_\ell$ elements to be zero. Due to the lack of available analytical form
for the posterior distribution of~$\gamma$'s, we employ the MCMC sampling
algorithm to sample from the posterior distribution, and then obtain the
posterior estimates of the unknown parameters. Computation is facilitated by the \textbf{nimble} \citep{de2017programming} package in \textsf{R}
\citep{Rlanguage2013}.

To determine the final clustering configuration based on post-burn-in
iterations, we use the Dahl's method \citep{dahl2006model}. The main idea is to
obtain a clustering configuration that best represents the posterior samples based on comparing the ``pairwise
similarity'' between different cluster structures. The procedure can be described as follows. First, at MCMC iteration~$t$, based on the
$n$-dimensional vector $(z_1^{(t)},\ldots,z_n^{(t)})$ for the latent clusters, a
membership matrix~$M^{(t)}$ consisting of~0 and~1's can be obtained, where 
$M^{(t)}(i,j) = M^{(t)}(j,i) = 1(z_i^{(t)} = z_j^{(t)})$. Next, the membership
matrices are averaged over all post-burn-in iterations to get a matrix of
pairwise similarity, $\bar{M} = \sum_{t=1}^T M^{(t)} / T$, where~$T$ denotes the
total number of iterations. Finally, the iteration that has the smallest
element-wise Euclidean distance from $\bar{M}$ is taken as the inferred
clustering configuration, i.e., with~$t^*$ being
\begin{equation*}
    t^* = \argmin_t \sum_{i=1}^n \sum_{j=1}^n \left(
    M^{(t)}(i,j) - \bar{M}(i,j)
    \right)^2,
\end{equation*}
and the final inferred configuration is obtaind as $(z_1^{t^*},\ldots, z_n^{t^*})$.

\section{Simulation}\label{sec:simu}
To evaluate the performance of the proposed model, simulation studies are
performed on generated data sets with a total of 150 players. We consider two simulation settings. For the first setting, we consider a three-angle pattern and three-distance partition of the court, i.e., the court is divided into $9$ parts based on combinations of distance and angle. Two clusters of size 75 are set for angle, distance, and quarter, respectively. The patterns for angle and group are visualized in Figure~\ref{fig:simu_pars_1} of the Supplementary Material. For
quarter group~1, we choose $\bm{\gamma}_3 = (-1, -1, -1, -1)^\top$; and for quarter group~2,
$\bm{\gamma}_3 = (-0.5, -2, -0.5, -2)^\top$. For the second simulation setting, we consider a finer partition of the court, including 11
angles, 12 distances, and again two quarter patterns in the same way as design~1.
The true number of clusters is 3 (each cluster cluster of size 50) for the angle, 3 (each cluster of size 50) for the distance, and 2 (each of size 75) for
the quarter. The angle and group patterns are visualized in Figure~\ref{fig:simu_pars_2} of the Supplementary Material. Under both settings, for each piece of the partitioned court, the corresponding
number of shots is generated using the associated $\gamma_1, \gamma_2$
and~$\gamma_3$ based on the last line of Equation~\ref{eq:hierarchical_model}.
The proposed multidimensional clustering approach
is then applied to fit the generated data; and this procedure is repeated for~100 times for each setting. All the computations were performed on a computing server (256GB RAM, with 8 AMD Opteron 6276 processors, operating at 2.3 GHz, with 8 processing cores in each) and the running time was within twelve hours.

\begin{figure}[tbp]
    \centering
    \includegraphics[width=0.8\textwidth]{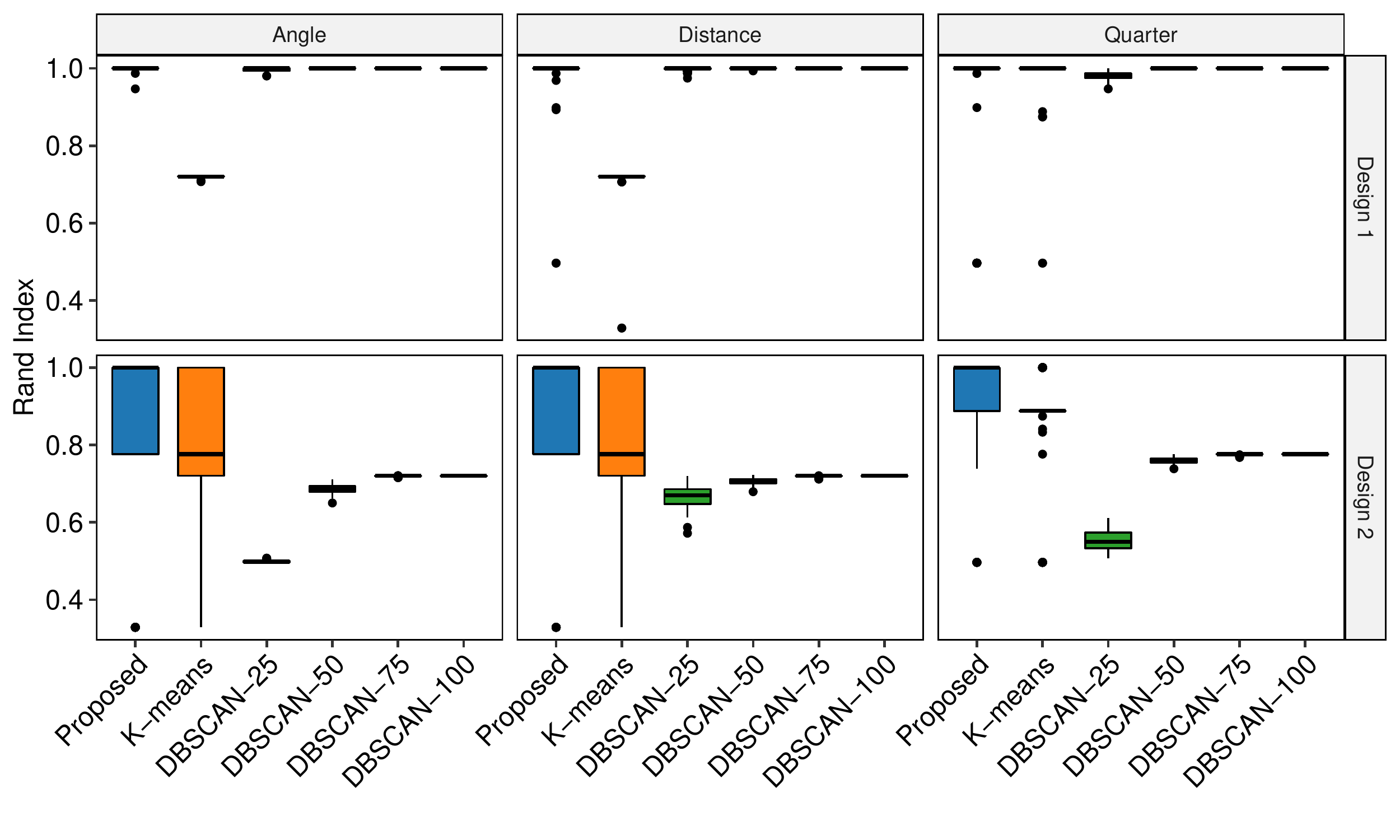}
    \caption{Simulation results: Rand index boxplots for angle, distance, and quarter over 100 Monte-Carlo replicates.}
    \label{fig:randindex}
\end{figure}

To evaluate the clustering performance on each of the tensor directions, we use the Rand index \citep[RI;][]{rand1971objective}, which is a commonly
used metric that measures the concordance between two clustering schemes. Taking values between~$0$ and~$1$, a larger value of RI indicates a higher agreement. To evaluate whether the true number of clusters is correctly inferred, we also examine the
total number of clusters inferred in each replicate over each of the three directions.

We consider two competing
methods,
$K$-means (function \texttt{kmeans()} in R) and density-based spatial
clustering \citep[DBSCAN; implemented in \textbf{fpc},][]{fpc2020}. To make a
fair
comparison, we use the number of clusters obtained by our method for
$K$-means. For DBSCAN, as the method depends on a
pre-specified ``reachability distance'', we use four candidate values, 25, 50,
75,
and 100; and we denote the methods as DBSCAN-25,$\ldots,$ DBSCAN-100 for the rest of
this paper. Both methods are applied to each of the three directions in an independent manner. In other words, for~150 simulated players, we sum out the
distance and quarter directions, and obtain~150 11-dimensional count vectors for clustering. 

We summarize the rand indexes from 100 replicates as boxplots in Figure~\ref{fig:randindex} and also report the average RI in
Table~\ref{tab:average_ri}. From the results, we find a clear advantage of our method over $K$-means. Compared to DBSCAN, our advantage is not obvious under the simple setting, Design 1; but becomes significantly better under Design 2. We also note that the performance of DBSCAN is quite sensitive to the choice of the reachability distance, e.g., DBSCAN-25 has the worst performance for all three directions under Design 2, but not for Design 1. Our method, on the other hand, manages to achieve a reasonably high average Rand Index for different tensor directions under both simulation designs. These results highlight the benefit of incorporating the tensor structure and borrowing information from other directions by our method. 

\begin{figure}[tbp]
    \centering
    \includegraphics[width=.7\textwidth]{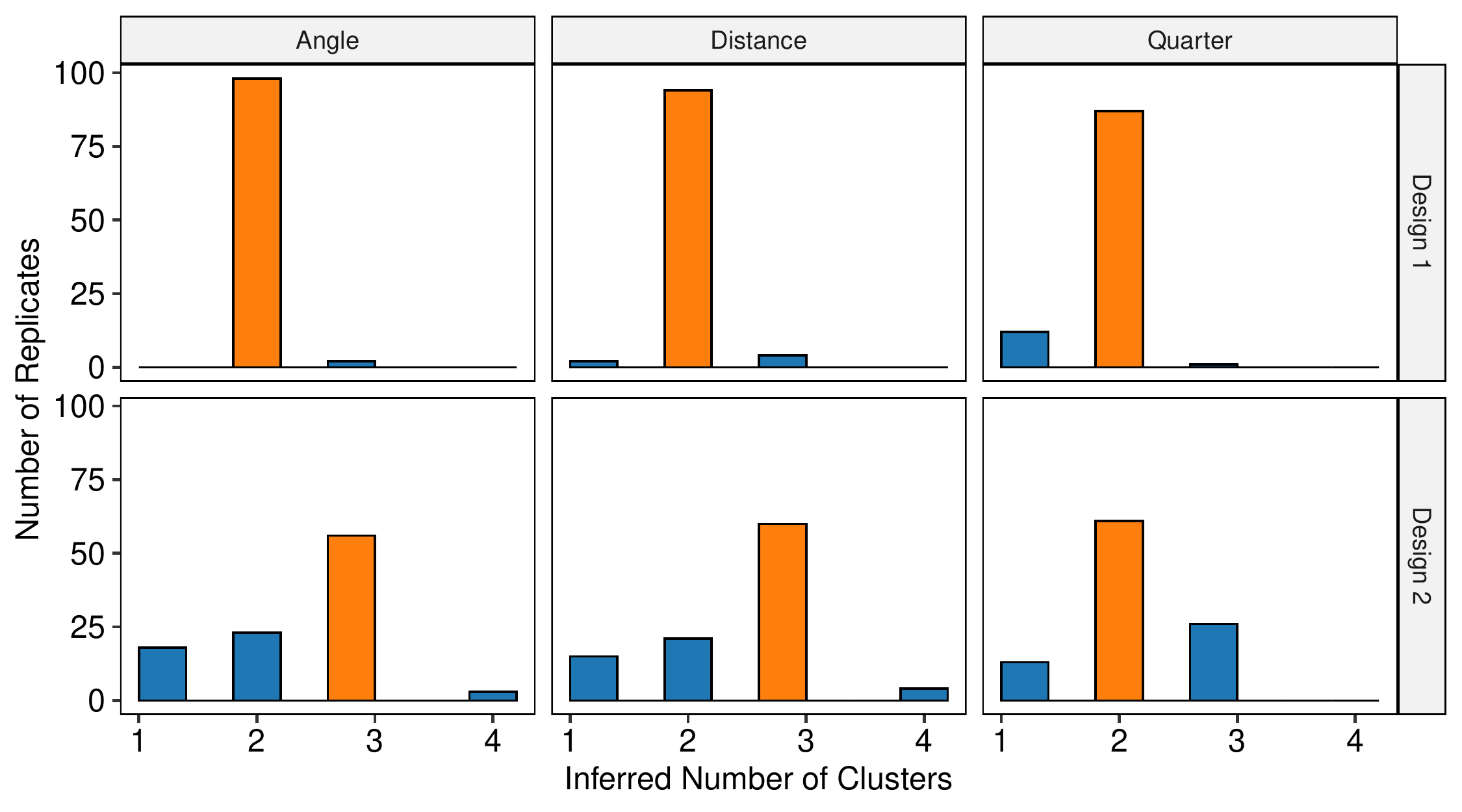}
    \caption{Simulation results: Histograms of cluster numbers over 100 Monte-Carlo replicates for angle,
distance, and quarter under two designs. Correct estimate of $K$ is marked in orange color. }
    \label{fig:K_hist}
\end{figure}

\begin{table}[tbp]
\centering 
\caption{Simulation results: Average Rand Index over 100 Monte-Carlo replicates under two simulation designs for the
proposed method and two competing methods.}
\label{tab:average_ri}
\begin{tabular}{llccc}
\toprule
Design &      Method &  Angle &  Distance &  Quarter \\
\midrule
    Design~1 &    Proposed &  0.999 &     0.987 &    0.938 \\ 
     &     $K$-means &  0.720 &     0.712 &    0.985 \\
     &   DBSCAN-25 &  0.996 &     0.998 &    0.982 \\
     &   DBSCAN-50 &  1.000 &     1.000 &    1.000 \\
     &   DBSCAN-75 &  1.000 &     1.000 &    1.000 \\
     &  DBSCAN-100 &  1.000 &     1.000 &    1.000 \\
\midrule
     Design~2 &    Proposed &  0.826 &     0.851 &    0.903 \\
     &     $K$-means &  0.773 &     0.786 &    0.851 \\
     &   DBSCAN-25 &  0.499 &     0.667 &    0.554 \\
     &   DBSCAN-50 &  0.687 &     0.706 &    0.759 \\
     &   DBSCAN-75 &  0.720 &     0.720 &    0.776 \\
     &  DBSCAN-100 &  0.720 &     0.720 &    0.776 \\
\bottomrule
\end{tabular}
\end{table}

We also present the histogram of the estimated number of clusters from 100 replicates in
Figure~\ref{fig:K_hist}. Under Design~1, where there are only two clusters on
each direction, the proposed method manages to correctly estimate the cluster number $98\%, 94\%$, and $87\%$ of times for
angle, distance, and quarter, respectively.
In Design~2, with a finer partition of the court, it becomes harder to infer
the number of clusters, and the percentage of correct estimation reduces to $56\%, 60\%$, and $61\%$.

\section{NBA Data Analysis}\label{sec:real_data}
We consider the shot attempts made by players during the 2017--2018 NBA regular
season excluding the overtime period. Rookie year players who started their NBA career in 2017 are excluded. We also exclude players who made very few number of shots in that season, e.g., due to long-term injury. Shots that were made at negative
degrees (under the polar coordinate system) are
also excluded. At the end, the dataset that we study consists of 122,001 shot attempts made by~191
players, with Aron Baynes bottoming the list with~317, and Russell Westbrook
topping the list with~1356 shot attempts.

\begin{figure}[tbp]
\centering
\includegraphics[width=.9\textwidth]{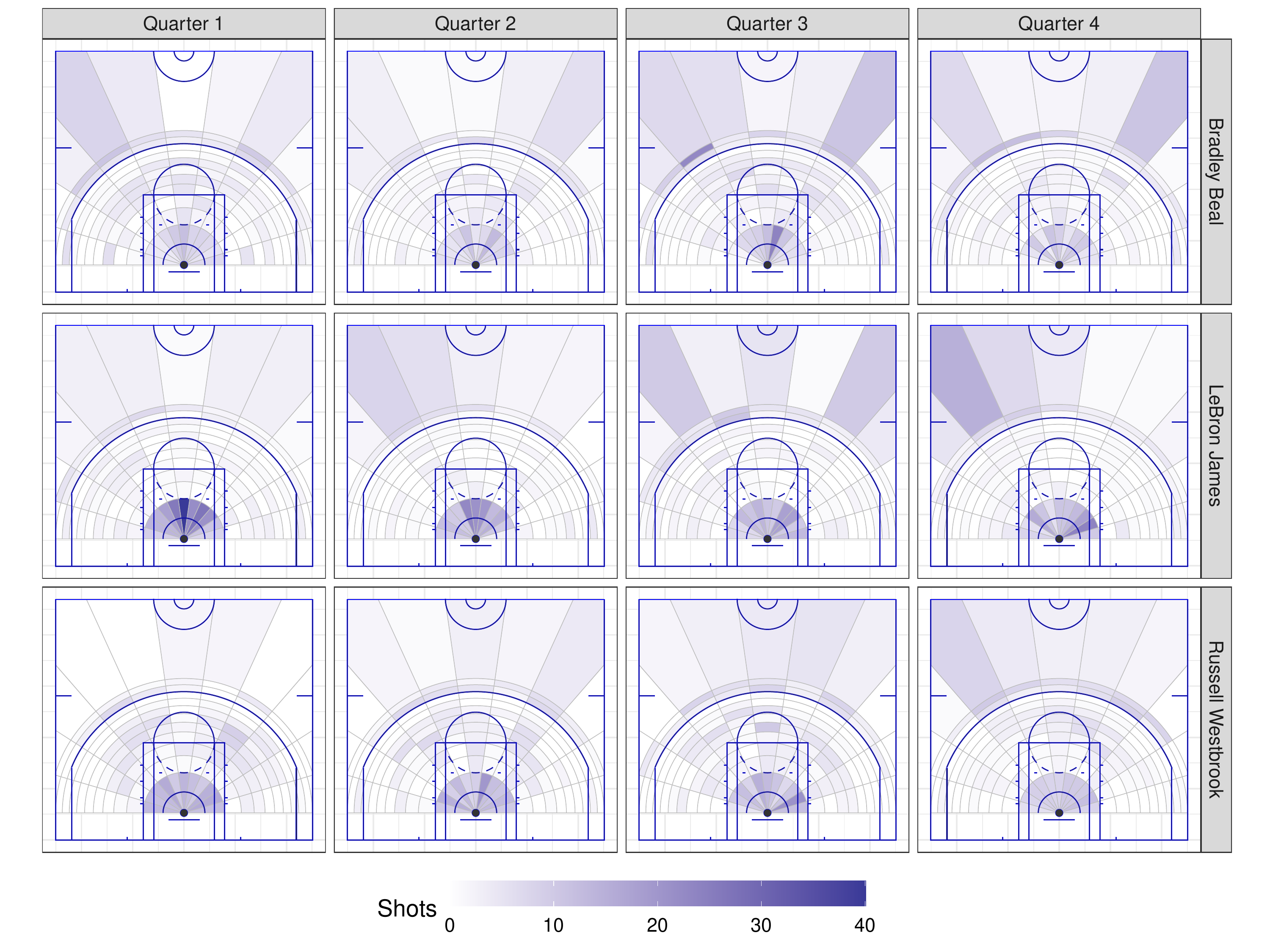}
\caption{Visualization of shot count tensors for Bradley Beal,
LeBron James, and Russel Westbrook.}
\label{fig:tensor_viz}
\end{figure}

We consider the polar coordinate representation of shot attempts in a similar way with
\cite{reich2006spatial}. We treat the basket as origin and partition the angle (from~0 to~$\pi$) into 11 equal sections. In
terms of the shooting distance, we partition it into 12 sections, with
the first~11 be designed so that the areas of all sectors and annular sectors
are the same. The remaining~9 areas correspond to the remaining areas on the
offensive half court. The partition scheme is illustrated in
Figure~\ref{fig:partition_scheme} of the Supplementary Material. Compared to the partition scheme in Figure~2
of \cite{reich2006spatial}, where the annular sectors only covered the
regions near the three-point line, we extend the annular sectors because of the
current trend of making three-point shots among NBA players, e.g., Stephen Curry
and Damian Lillard. For each player, we further divide the number of shot attempts by four game quarters for each court partition, and end up with a~$11\times 12 \times 4$-dimensional tensor. In Figure~\ref{fig:tensor_viz}, we  choose three players, Bradley Beal, LeBron James, and Russel Westbrook, and present their shot charts for demonstration. Some interesting patterns can be observed from the plots, e.g., LeBron James makes more shots facing the basket, and Russel Westbrook makes fewer shot attempts in the fourth quarter on average.


We apply the proposed multidimensional heterogeneity learning approach on the collected tensor data from 191 players. The same
neighborhood matrices $W_1$, $W_2$, and $W_3$ from the simulation studies are
used. We consider a MCMC chain of length 10,000 and  and a thinning interval of 2, resulting in a total of 5,000 posterior samples. We then discard the first 2,000 as burn-in and use the rest of 3,000 samples to obtain the final clustering configuration using Dahl's method as described in Section \ref{sec:inference}. 

\begin{figure}[tbp]
\centering
\includegraphics[width = 0.55\textwidth]{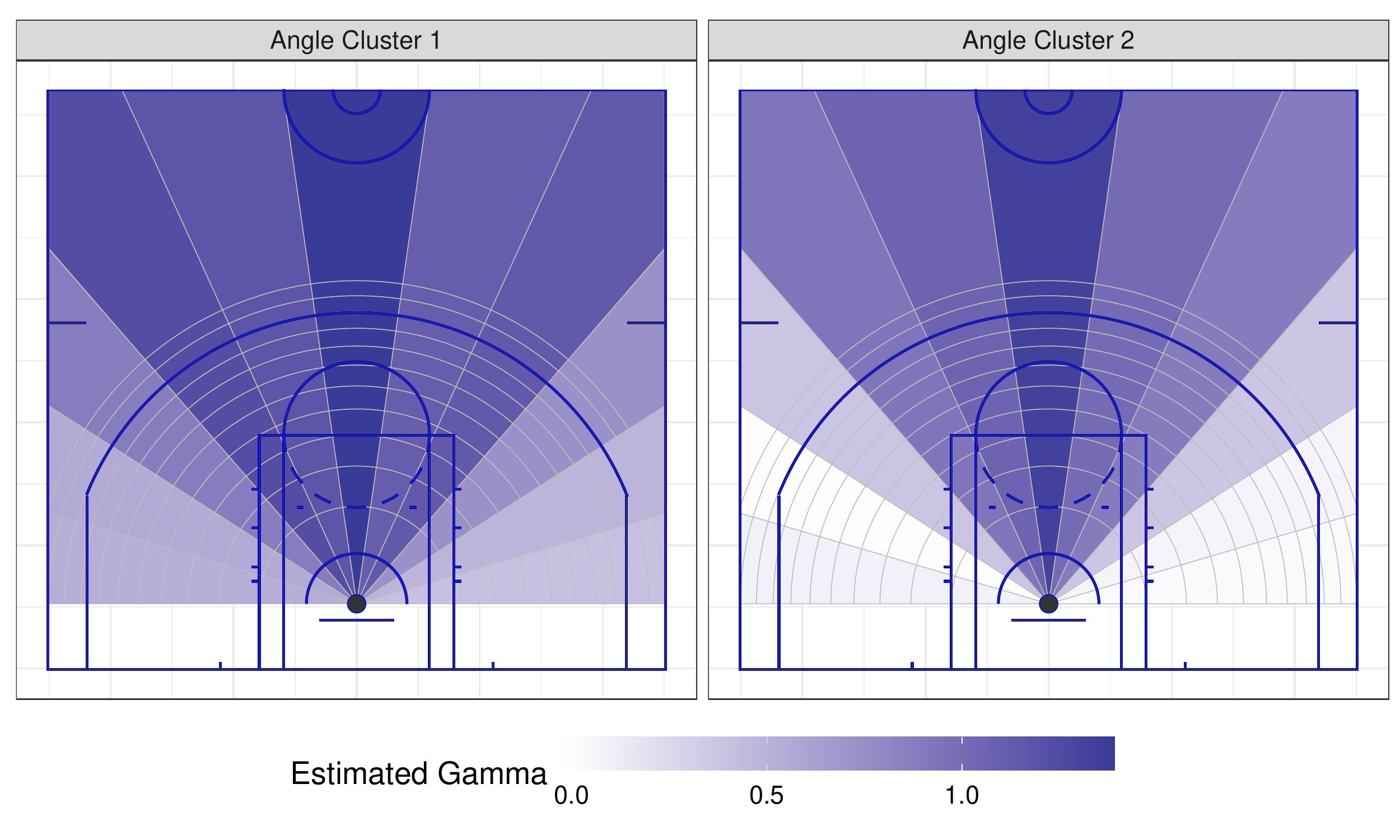}
\caption{Visualization of $\bm{\gamma}_1$ estimates for two shooting angle clusters.}
\label{fig:angle_cluster}
\end{figure}

We obtain two clusters of sizes 71 and 120 over the angle direction as shown in Figure~\ref{fig:angle_cluster}. While it can be seen that players in both clusters make more shots when facing the basket, those in
cluster~1 also make a fair amount of shots at the two wings, as well as the
corners. Players in cluster~2, however, mostly shoot in the region facing the
basket and its immediate neighbors. Compared to those in cluster~1, they make
less corner shots, as the estimated~$\bm{\gamma}_1$ is almost~0 in the two
regions on each side. Representative players for the two clusters are,
respectively, James Harden and John Wall. Their shot charts are given in Figure \ref{fig:angle_cluster_players} of the Supplementary Material. 

\begin{figure}[tbp]
\centering
\includegraphics[width = .7\textwidth]{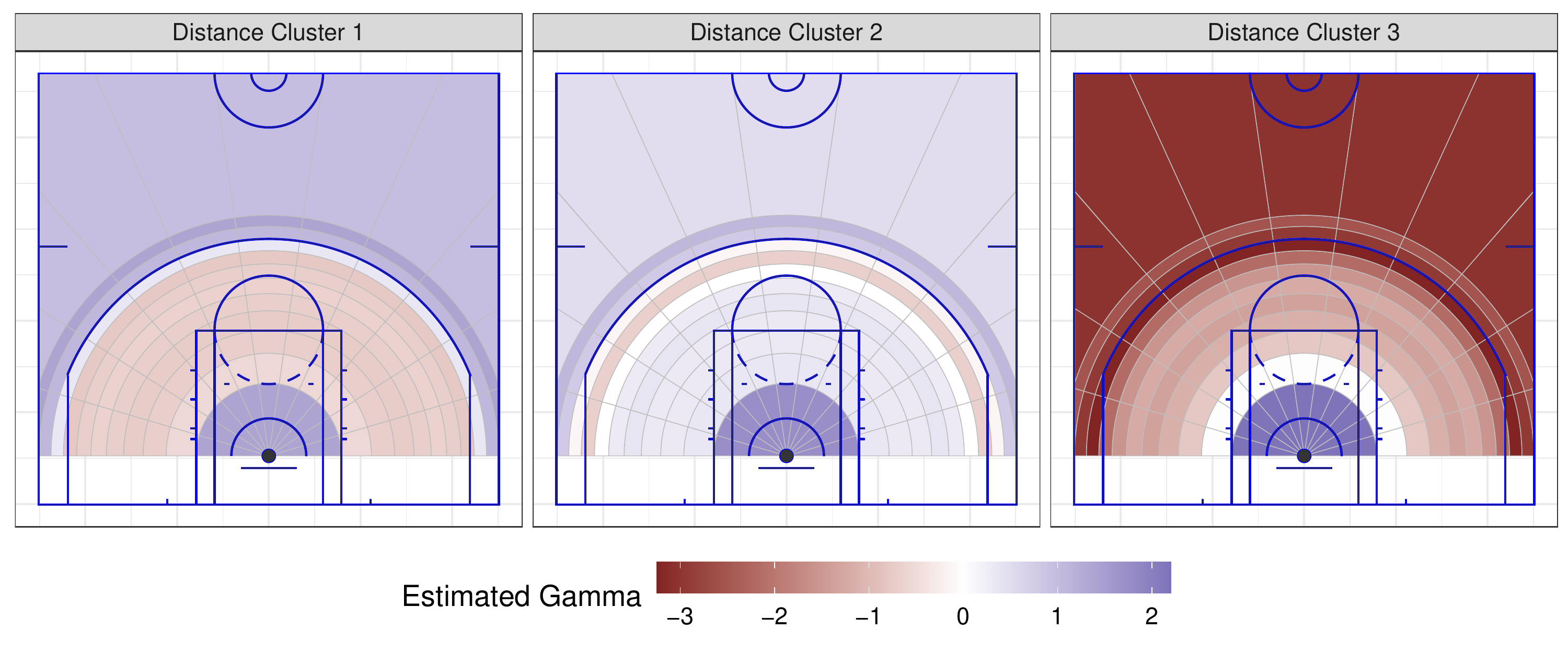}
\caption{Visualization of $\bm{\gamma}_2$ estimates for the three shooting distance clusters.}
\label{fig:dist_cluster}
\end{figure}

Shooting patterns in terms of distance to basket have been clustered into three
groups as visualized in Figure~\ref{fig:dist_cluster}. Players in the cluster~1
have two hot regions: near the basket, and beyond the three point line. Point
guards and shooting guards (small forwards) make the majority of this cluster (75 players), with representative players such as Kyrie Irving and Stephen Curry. Compared with cluster~1, the 90 players in cluster~2 tend to shoot less beyond the three
point line, but make more perimeter shots. A representative player for this
cluster is Russell Westbrook. Finally, in cluster~3, most of the 26 players only
shoot in regions that are closest to the basket, such as DeAndre Jordan and
Clint Capela. Most of their shots are slam dunks and alley-oops. Some other players (e.g., Fred VanVleet and Jonas Valanciunas) in
cluster~3, although also making perimeter shots and three-pointers, rely
heavily on lay-ups. We pick one representative player from each cluster and present their shooting charts in Figure \ref{fig:dist_cluster_players} of the Supplementary Material.

\begin{figure}[tbp]
\centering
\includegraphics[width = .8\textwidth]{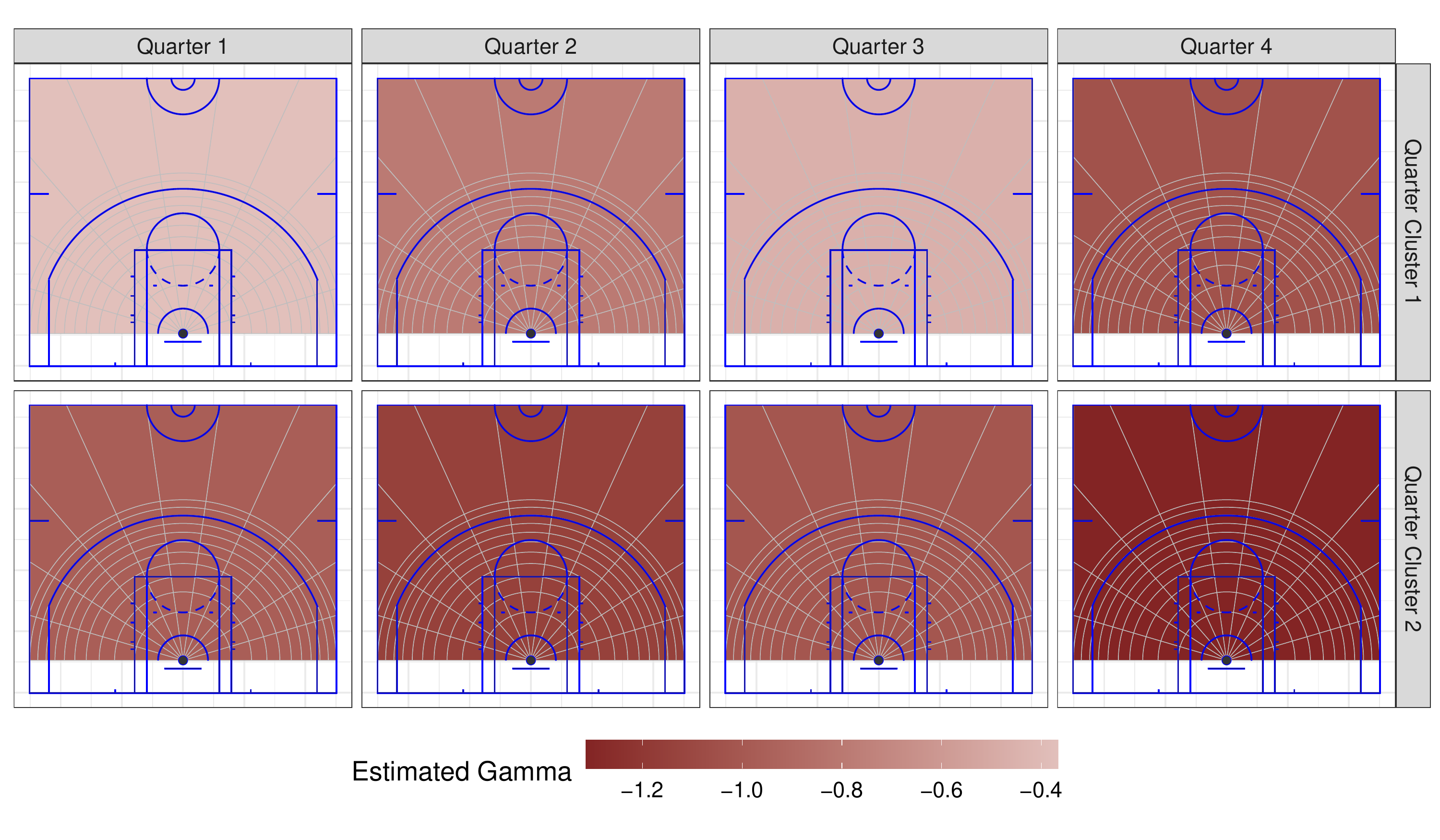}
\caption{Visualization of $\bm{\gamma}_3$ estimates for two game quarter clusters.}
\label{fig:quarter_cluster}
\end{figure}

Finally, the two clusters for quarters are visualized in Figure~\ref{fig:quarter_cluster}. In cluster~1, players make more shots in
quarters~1 and~3, and less shots in quarters~2 and~4. Most players in this
cluster are leading players in their teams, and they often take
breaks during the second quarter. In the fourth quarter, leading players may
also take breaks if their teams lead or fall behind by wide margins. Stephen
Curry, Kevin Durant and Paul George are in this cluster. In cluster~2, the
distribution of shots across four quarters is more even than that in cluster~1,
and on average the estimated $\bm{\gamma}_3$ is relatively smaller. The cluster
sizes are 91 and 100, which indicate these two patterns are similarly prevalent
among the players that we studied. We pick Anthony Davis and Chris Paul as two representative players from two clusters and present their shooting charts over four quarters in Figure \ref{fig:quarter_cluster_players} of the Supplementary Material.

\section{Discussion}\label{sec:discussion}
We propose a new multidimensional tensor clustering method in this paper and demonstrate its utility by studying how shooting patterns distribute over court locations and game time among different players. Our method is applicable to many other sports such as football and baseball, where it is natural to formulate and model the multi-way array data. The proposed method also applies to other applications such as imaging analysis and recommender systems. 

Several future work directions remain open. First, our method is based on Poisson distribution for outcomes in the tensor; and it is of interest to generalize this assumption by considering other types of distributions such as zero-inflated Poisson and continuous distributions. Incorporating sparsity in tensor models is another interesting direction that will allows us to deal with high-dimensional tensors. From applications point of view, it is of interest to analyze and compare the shooting patterns between different periods of games/seasons, e.g., regular season versus playoff games, and before-pandemic versus 2020 NBA bubble seasons.

\bibliography{tensor_cluster}
\bibliographystyle{chicago}


\newpage

\section*{Supplemental Material}
 
We present additional figures referenced in the main paper.

\begin{figure}[ht]
\centering
\includegraphics[width=0.5\textwidth]{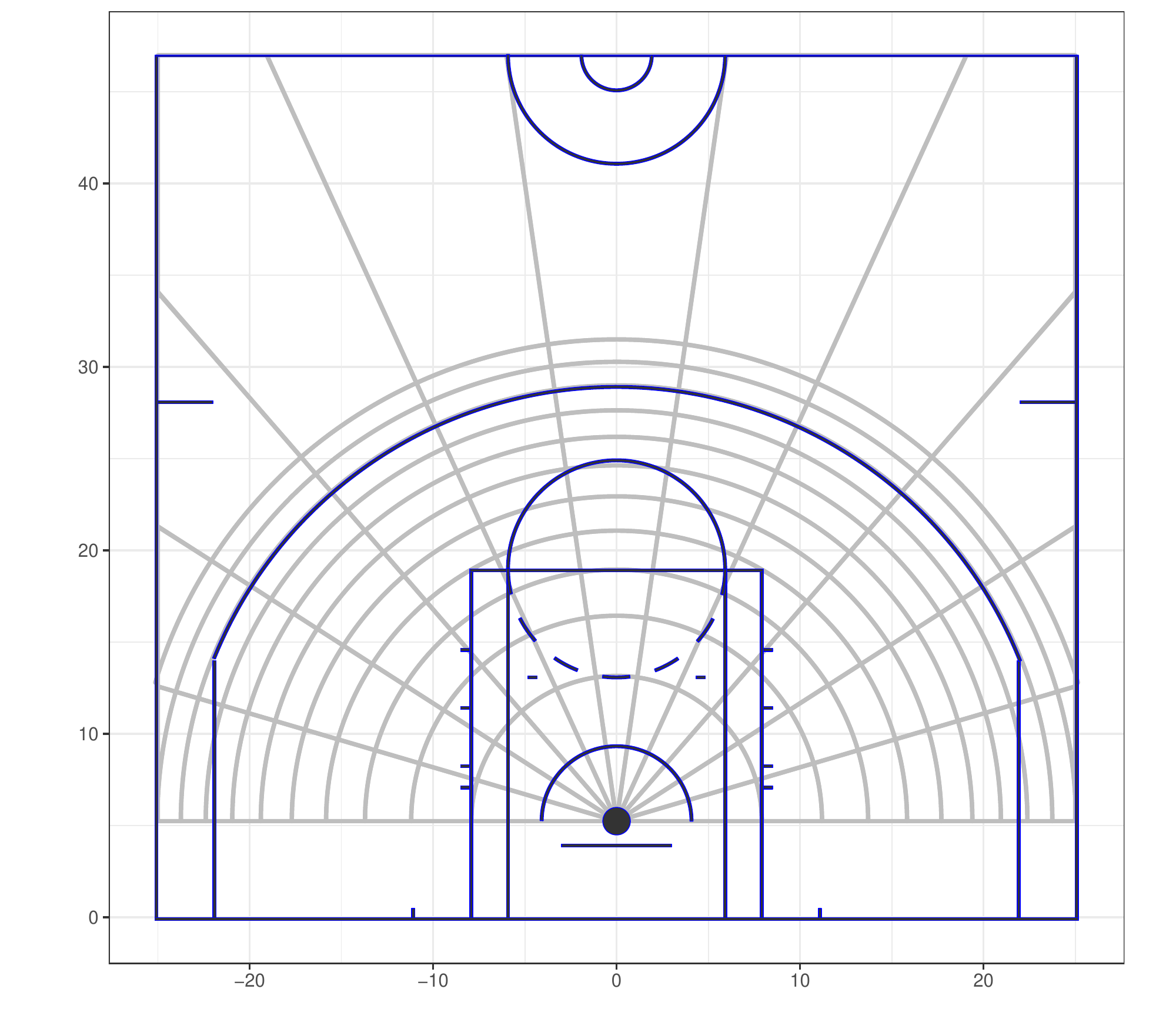}
\caption{Illustration of the partition scheme imposed on the court.}
\label{fig:partition_scheme}
\end{figure}

\begin{figure}[ht]
\centering
\includegraphics[width=0.5\textwidth]{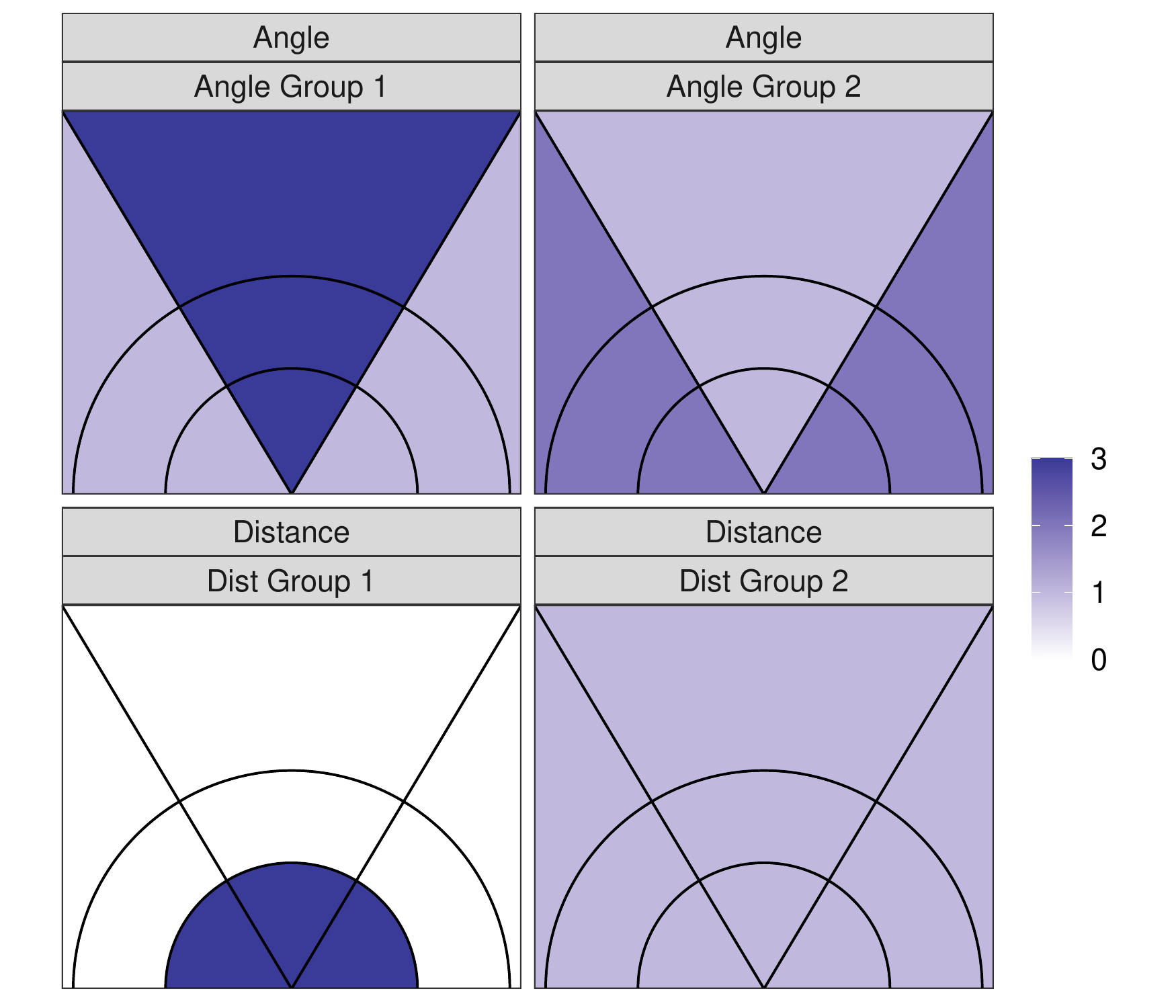}
\caption{Visualization for $\bm{\gamma}_1$ and $\bm{\gamma}_2$ in the first
simulation setting.}
\label{fig:simu_pars_1}
\end{figure}

\begin{figure}[ht]
\centering
\includegraphics[width = 0.75\textwidth]{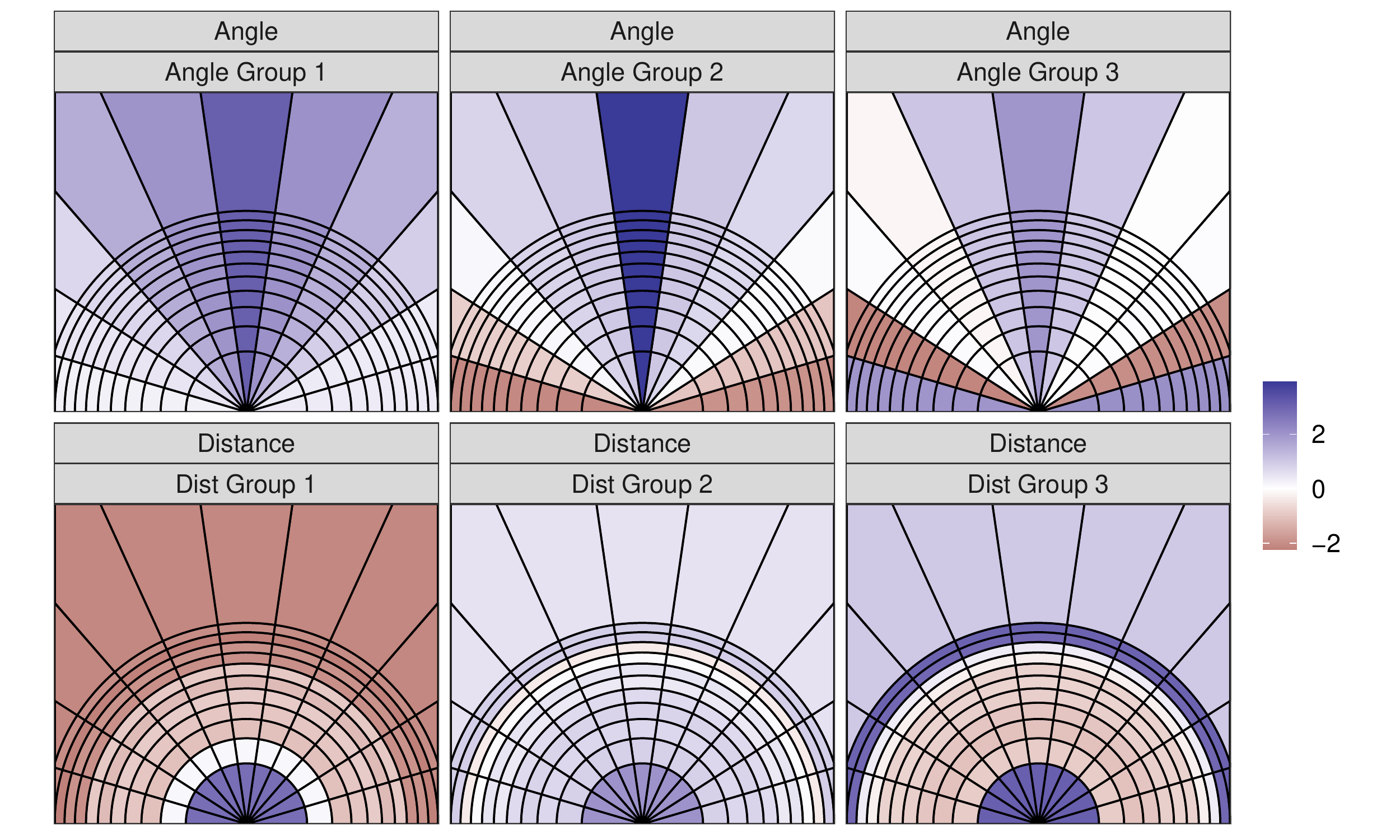}
\caption{Visualization for $\bm{\gamma}_1$ and $\bm{\gamma}_2$ in the second
simulation setting.}
\label{fig:simu_pars_2}
\end{figure}

\begin{figure}[tbp]
  \centering
  \includegraphics[width = 0.6\textwidth]{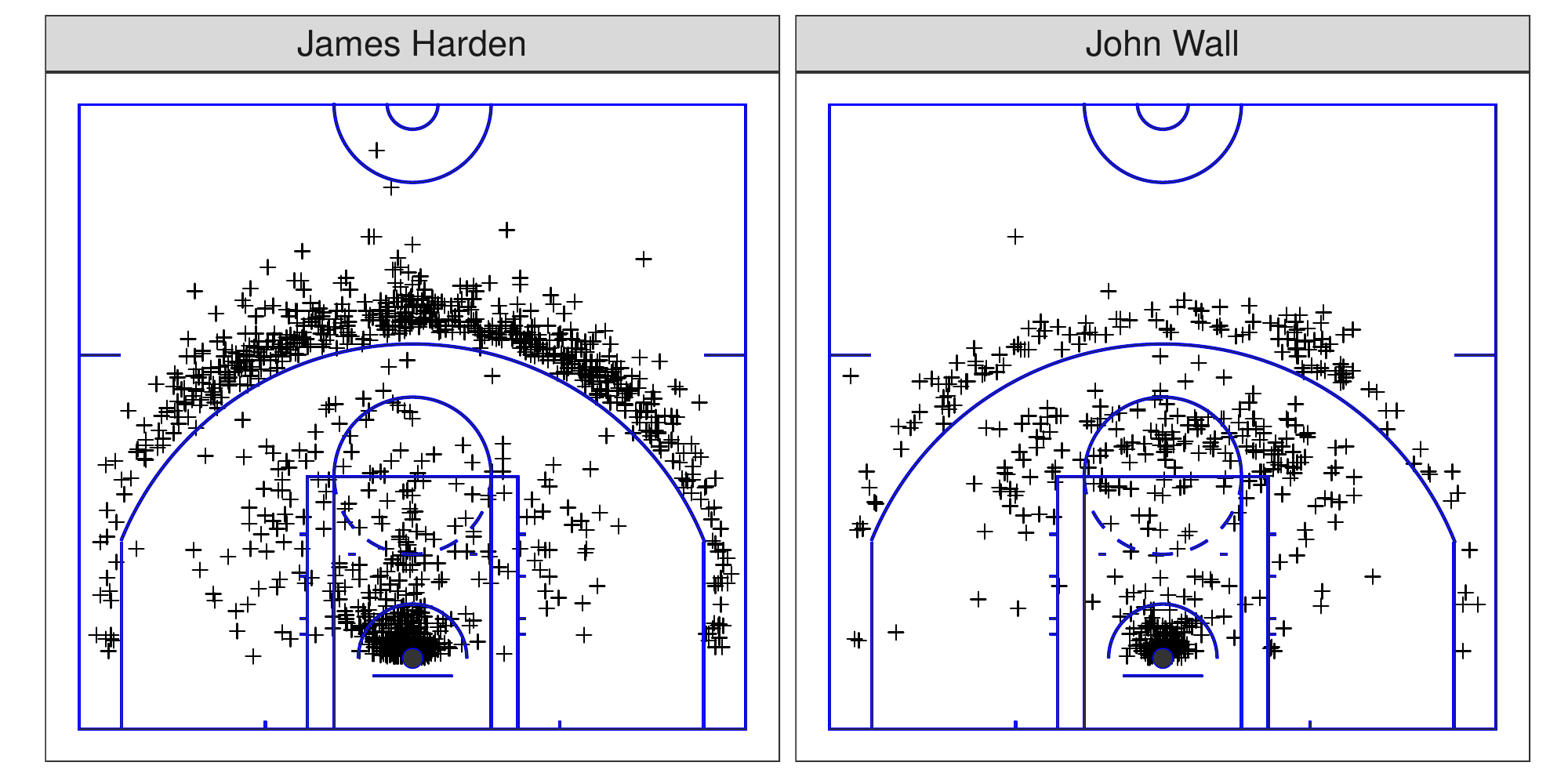}
  \caption{Shot charts for representative players from the two angle clusters.}
  \label{fig:angle_cluster_players}
\end{figure}

\begin{figure}[tbp]
  \centering
  \includegraphics[width = 0.9\textwidth]{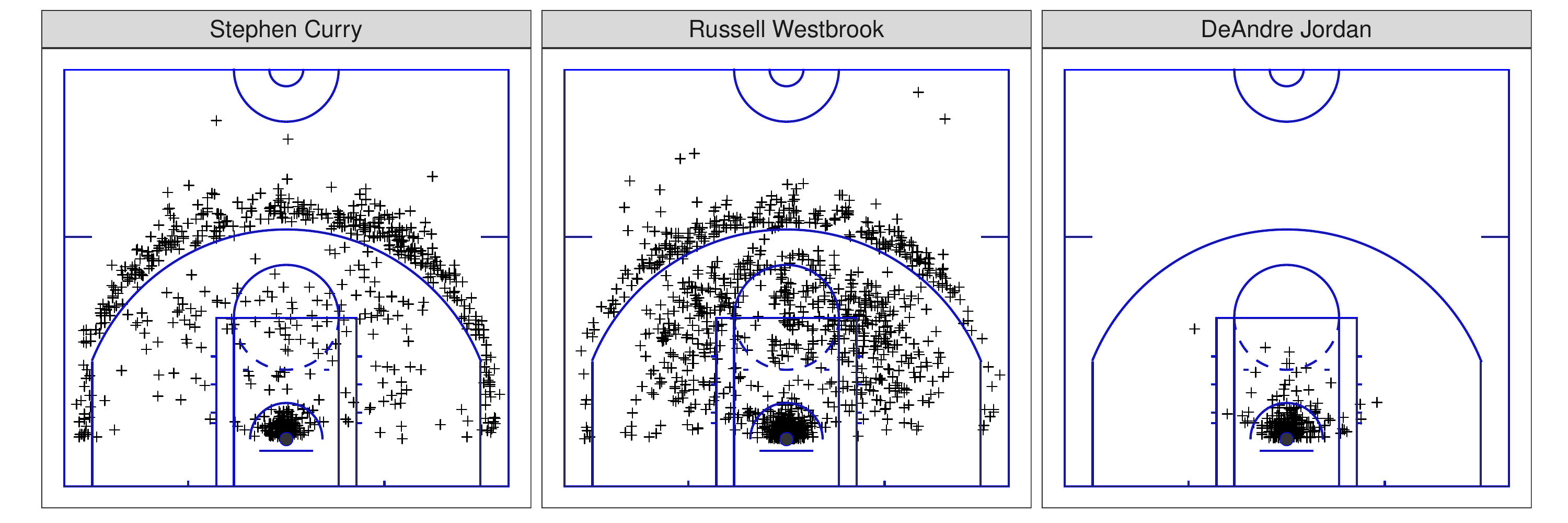}
\caption{Shot charts for representative players from the three distance
clusters.}
  \label{fig:dist_cluster_players}
\end{figure}

\begin{figure}[tbp]
  \centering
  \includegraphics[width = \textwidth]{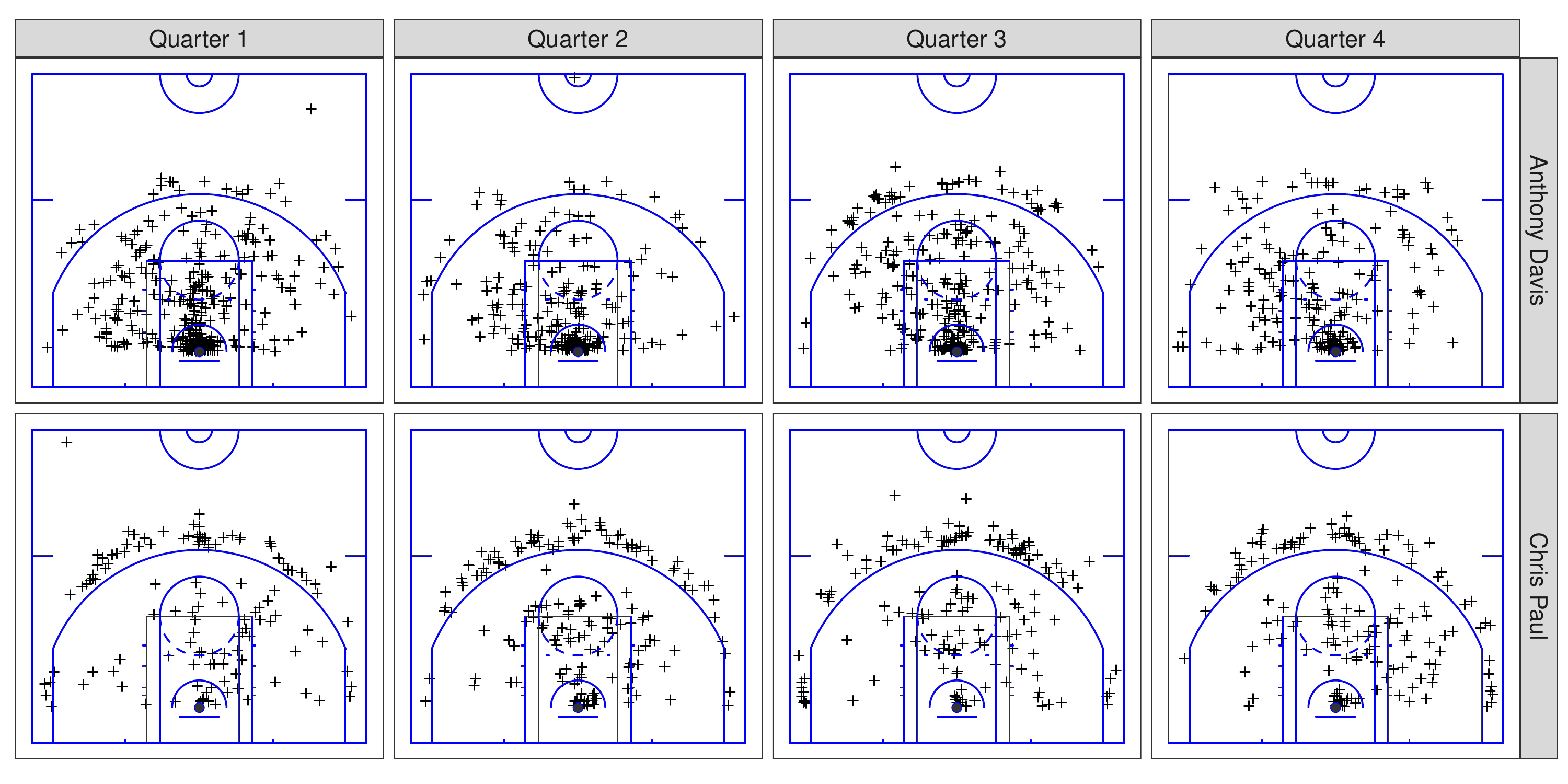}
\caption{Shot charts for representative players from the two quarter clusters.}
  \label{fig:quarter_cluster_players}
\end{figure}

\begin{figure}[tbp]
\centering
\includegraphics[width = 0.9\textwidth]{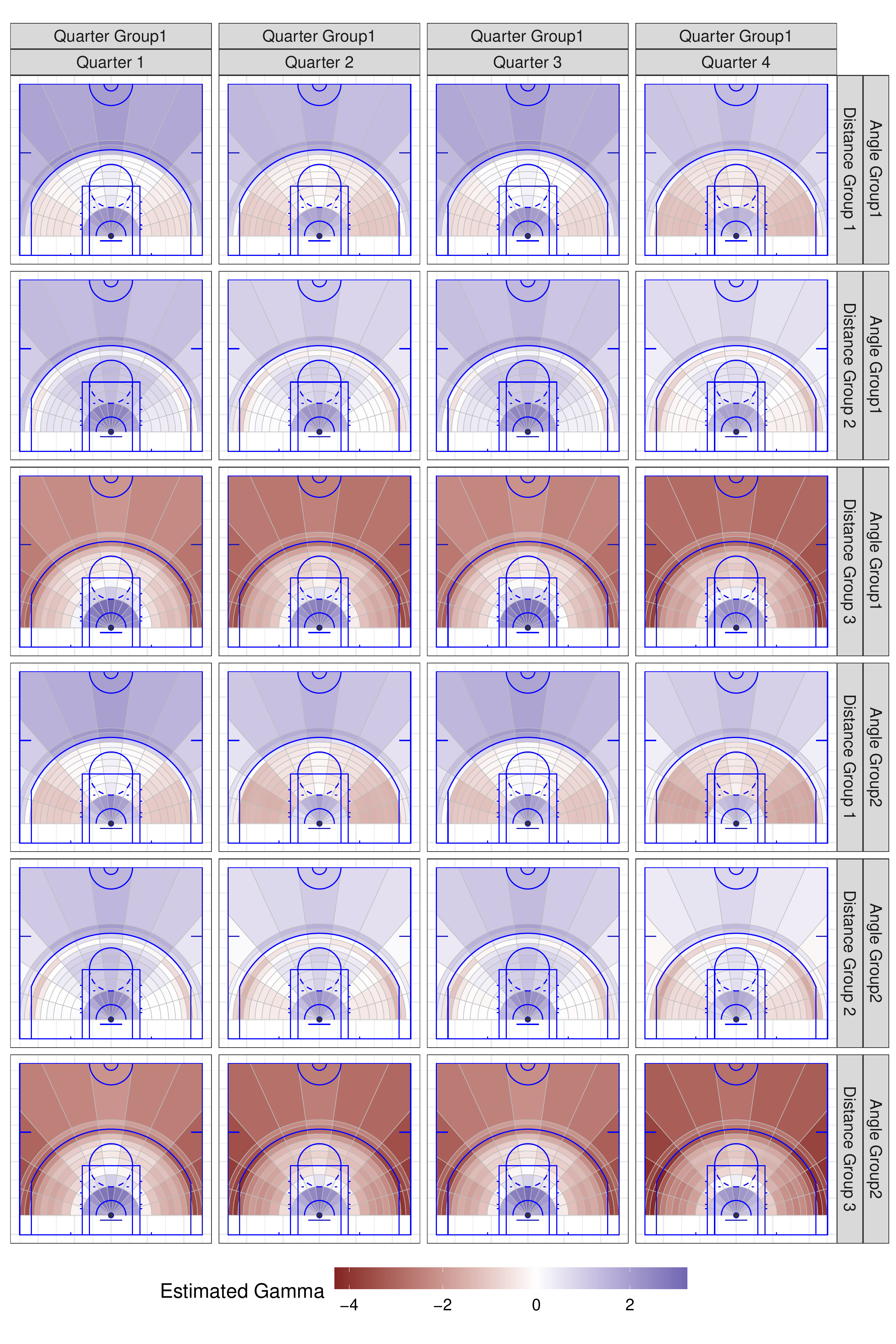}
\caption{Visualization of $\hat{\gamma}_1 + \hat{\gamma}_2 + \hat{\gamma}_3$
for players in different distance and angle groups and quarter group~1.}
\end{figure}

\begin{figure}[tbp]
\centering
\includegraphics[width = 0.9\textwidth]{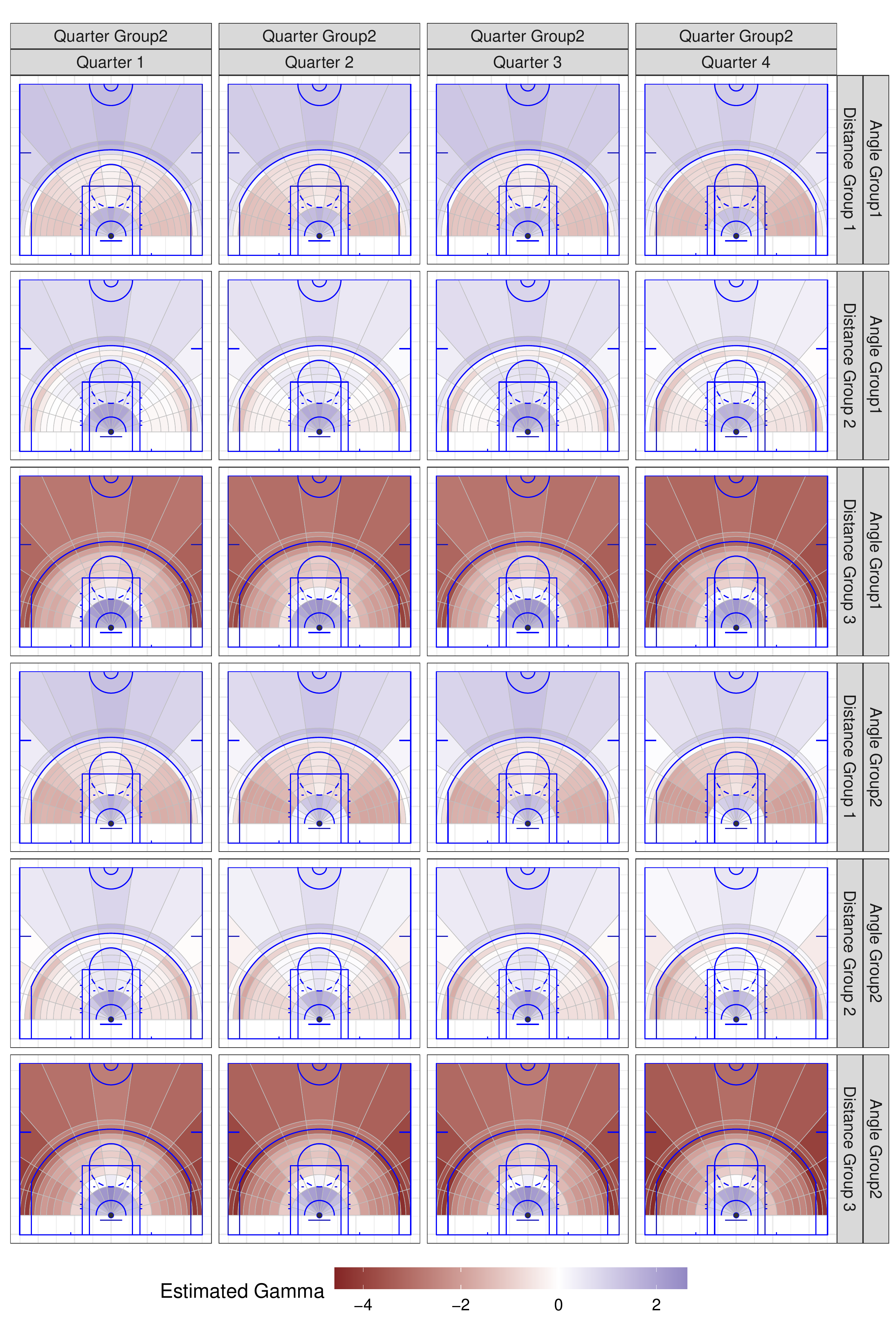}
\caption{Visualization of $\hat{\gamma}_1 + \hat{\gamma}_2 + \hat{\gamma}_3$
for players in different distance and angle groups and quarter group~2.}
\end{figure}

\end{document}